\newcommand{\cuo}{CuO$_2$}

\newcommand{\OA}{O$_{\rm A}$}
\newcommand{\OP}{O$_{\rm P}$}
\newcommand{\rme}{{\rm e}}
\newcommand{\tc}{$T_{\rm c}$}
\documentclass[superscriptaddress,prl,twocolumn,
amsmath,amssymb,showpacs]{revtex4}
\usepackage{graphicx}

\begin{document}

\title{Origin of the spatial variation of the pairing gap in Bi-based 
high temperature cuprate superconductors}

\author{Michiyasu Mori}
\affiliation{Institute for Materials Research, Tohoku University, Sendai 
980-8577, Japan}

\author{Giniyat Khaliullin}
\affiliation{Max-Planck-Institut f\"ur Festk\"orperforschung,
Heisenbergstrasse 1, D-70569 Stuttgart, Germany} 

\author{Takami Tohyama}
\affiliation{Yukawa Institute for Theoretical Physics, Kyoto University, 
Kyoto 606-8502, Japan}

\author{Sadamichi Maekawa}
\affiliation{Institute for Materials Research, Tohoku University, Sendai
980-8577, Japan} 
\affiliation{CREST, Japan Science and Technology Agency, 
Kawaguchi 433-0012, Japan}

\begin{abstract}
Recently, scanning tunneling microscopy on Bi-2212 cuprate 
superconductor has revealed a spatial variation of the energy 
gap that is directly correlated with a modulation of the apical 
oxygen position. We identify two mechanisms by which out-of-plane 
oxygens can modulate the pairing interaction within the CuO$_2$ 
layer: a covalency between the $x^2$-$y^2$ band and apical $p$-orbital, 
and a screening of correlation $U$ by apical oxygen polarization. Both  
effects strongly depend on the apical oxygen position and their cooperative
action explains the experiment. 
\end{abstract}

\date{\today}


% 74.72.Hs Bi-based cuprates
% 74.62.Bf Effects of material synthesis, crystal structure, 
%and chemical composition
% 74.20.-z Theories and models of superconducting state
% 75.30.Et Exchange and superexchange interactions

\maketitle

Various types of high-\tc~cuprates have been discovered for the last 
few decades. As a function of density of charge carriers doped into 
the \cuo~planes, a superconducting critical temperature \tc~shows 
in general a maximum which varies strongly from one family of cuprates
to another. For hole-doped cuprates, a correlation between the maximum \tc~and 
the energy-level separation of in-plane oxygens (\OP) and apical 
ones (\OA) has been noticed early on \cite{Oht91,Mae04}. Hence, a role 
of apical oxygen on \tc~is of considerable 
interest \cite{Oht91,Mae04,Mat90,Pav01,Eis06,Liu06,Mc05}.  

In Bi$_2$Sr$_2$CaCu$_2$O$_8$ (Bi-2212), a mismatch between the rock-salt 
BiO$_2$ layers and the CuO$_2$ planes causes an extra modulation of the 
crystal structure with a period about 26 \AA. In such a "supermodulated" 
lattice, the distance $d$ from the CuO$_2$ layer to the 
apical oxygen \OA~is periodically varied within the range 
of $\sim \pm$6\% \cite{Yam90}. Given that bonds 
within the CuO$_2$ plane itself are much less affected by supermodulation, 
Bi-2212 material provides a unique opportunity to study the impact of apical 
oxygen on superconductivity, by monitoring local electronic properties as 
a function of $d$ that varies spatially within a supermodulation period. This 
is precisely what is done in the recent scanning tunneling microscopy (STM) 
experiment by Slezak {\it et al.} \cite{Sle07,Sle08} 
(see also Ref.~\onlinecite{Nor08}). 

Slezak {\it et al.} measured the gap $\Delta$ in a local density of electronic
states, and found nearly 10\% spatial variation of $\Delta$ with 
the same periodicity as supermodulation. They have emphasized that 
the gap variation is {\it anticorrelated} with the Cu-\OA~distance 
variation, $\delta\Delta \propto -\delta d$, i.e., the gap increases 
when \OA~gets closer to the CuO$_2$ plane and vice versa. 

This remarkable observation has already been addressed in several papers 
by introducing variations of coupling constants on a phenomenological 
level \cite{And07,Yan07}. However, the underlying microscopic mechanism 
that links the strength of pairing interactions {\it within} CuO$_2$ planes 
with the position of {\it out-of-plane} \OA~remains elusive. 
In this Letter, we discuss the physical origin of high sensitivity 
of the pairing gap to the Cu-\OA~distance and explain the 
anticorrelation effect $\delta\Delta \propto -\delta d$ observed.  

Quite in general, the structural shift of apical oxygen may influence 
the energy gap either via the hopping parameters (hence density of states 
on the Fermi level), or via the strength of the pairing interaction. 
In particular, there is a well-known relation between the 
next-nearest-neighbor hopping $t$' and \tc, based on the band structure 
calculations by Pavarini {\it et al.} \cite{Pav01}. One has to notice, 
however, that this observation concerns a comparison between 
{\it different families of cuprates} with different lattice structure. 
In fact, Pavarini {\it et al.} predicted that the variation of the Cu-\OA~ 
distance {\it within a given compound} hardly affects the hopping 
parameters \cite{lines}. Therefore, we focus here on possible electronic 
mechanisms by which apical oxygens may affect the strength 
of the pairing potential within the $x^2$-$y^2$ band \cite{phonon}.  

We find two different ways how the apical oxygen may enter the game.  
First, a hybridization of "useful" $x^2$-$y^2$ band with the "pairing-inert" 
orbitals of apical oxygens reduces the pairing interaction. Such a destructive 
effect of covalency is controlled by a relative energy separation between the 
orbital levels that depends on Cu-\OA~distance. We illustrate this 
by an explicit calculation of the Madelung potential as functions of $d$. 
Second, we show that the superexchange interaction $J$,  
which is believed to be essential for magnetic correlations and 
possibly for superconductivity, is very sensitive to the apical \OA~
position. This is due to the high polarizability of O$^{2-}$ anion which has 
the effect of screening and reducing the energy $U$ needed to move an 
electron from one ion to another \cite{Boe84,Bri95}. In Bi-2212, 
the screening effect and hence the strength of magnetic correlations 
$J\propto 1/U$ are spatially modulated because the closer the apical 
\OA~is, the stronger the screening is. Remarkably, we find that the 
above two effects, covalency and screening, both favor an anti-phase 
relation between $\Delta$ and $d$ variations. 

{\it Covalency.}-- We address this effect in terms of the following
Hamiltonian:  
\begin{eqnarray}
H&=&\sum_{k\sigma}\epsilon_k c^\dag_{k\sigma}c_{k\sigma}
-\sum_k \Delta(\gamma_k^d c^\dag_{k\uparrow}c^\dag_{-k\downarrow} + h.c.) 
\nonumber\\
&+&\sum_{k\sigma}\epsilon_{\rm A} a^\dag_{k\sigma}a_{k\sigma}  
+\sum_{k\sigma}\nu_{k} (a^\dag_{k\sigma}c_{k\sigma}+h.c.). 
\label{hamil}
\end{eqnarray}
Here, the first term corresponds to the in-plane $pd\sigma$-band of 
$x^{\rm 2}$-$y^{\rm 2}$ symmetry made of Cu and \OP~orbitals, and a
conventional form of dispersion \cite{Pav01}  
$\epsilon_k=-2t(\cos k_x+\cos k_y)
+4t'\cos k_x \cos k_y-2t''(\cos 2k_x+\cos 2k_y)-\mu$, 
where $\mu$ is the chemical potential, 
is adopted. 
Second term shows that this band is supposed to host a 
superconductivity of $d$-wave symmetry, with the gap function 
$\gamma_k^d=(\cos k_x-\cos k_y)/2$ and the gap magnitude 
$\Delta= g\sum_{k}\gamma_k^d\langle c_{-k\downarrow}c_{k\uparrow}\rangle$, 
determined by the strength $g$ of a pairing potential (whose origin 
is not specified). 

Third term in Eq.~(\ref{hamil}) represents holes on the 2$p_z$ orbital of 
apical \OA~with energy $\epsilon_{\rm A}$. 
While a complete model may consider the rich internal structure of 
the \lq\lq axial\rq\rq~orbital by Pavarini {\it et al.} including 
3$d_{3z^2-r^2}$ and 4$s$ states of Cu \cite{Pav01}, we consider 
here a minimal model that captures the essential effects of the axial 
orbitals. Finally, the last term accounts for a covalent mixing of 
the $x^2$-$y^2$ and $p_z$ orbitals. $d$-wave symmetry of the matrix
element $\nu_k=4\nu\gamma_k^d$ is imposed by the hopping geometry: 
a transfer integral $\nu$ from $x^2$-$y^2$ type Zhang-Rice orbital to the  
neighboring $p_z$ states of axial symmetry must have different 
signs along $x$ and $y$ directions. 

Physically, we associate $\epsilon_{\rm A}$ with the energy-level 
separation of holes residing on in-plane \OP~and apical \OA~sites, 
as shown schematically in Fig.~1(b). A magnitude of $\epsilon_{\rm A}$ 
can then be estimated from the Madelung potentials on \OP~and 
\OA~\cite{note1} using the structural data of Ref.~\onlinecite{Yam90}. 
To calculate variations of $\epsilon_{\rm A}$ caused by supermodulation, 
we use the displacement pattern, Fig.~1(a), inferred from the structural 
data. This way, we quantify $\epsilon_{\rm A}$ in terms of the distance 
$d$ from the top CuO$_2$ layer (relevant for STM) to \OA~above it, making
thereby a link between the model and supermodulation. 

Below, we use a representative hopping parameters $t$=0.4 eV, $t'/t$=0.3, 
$t''=t'/2$, and $\nu/t=0.35$. The $d$-wave momentum dependence of 
$\nu_k$ renormalizes the $t',t''$ values but we compensate this 
numerically by adding a counter-term $\propto \nu^2/\epsilon_{\rm A}$ 
and keep the actual values of $t',t''$ invariant against the \OA~shifts 
(as found in the band structure calculations \cite{Pav01}). 
The covalency effect is then entirely due to 
the spectral weight shifts between $x^2$-$y^2$ and axial orbitals. 

%%%%%%%%%%%%%%%
\begin{figure}[t]
\includegraphics[height=6cm]{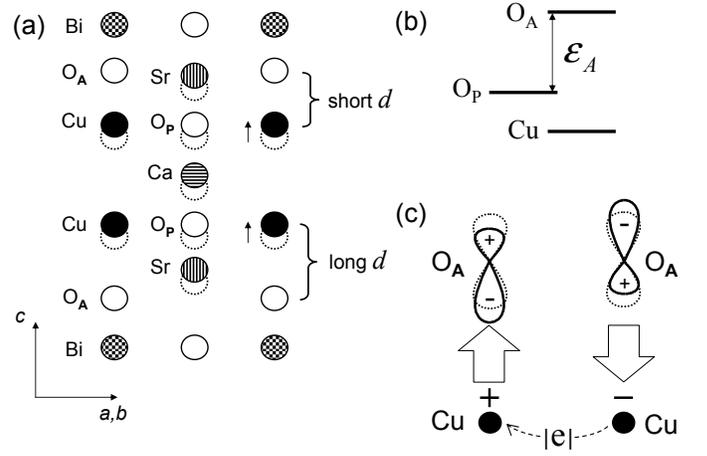}
\caption{
(a) Local atomic displacements along the $c$ axis due to the 
supermodulation (exaggerated). 
The Bi-\OA~bond length does not change much but other atoms 
(Cu, \OP, Sr, and Ca) are shifted such that the Cu-\OA~bond is most 
affected \cite{Yam90}. The direction (up/down) and amplitude of the shift 
$\delta d$ modulates along the $a$ axis with a period $\sim$26 \AA, 
resulting in a $\delta d \sim \pm$6\%~variation of the Cu-\OA~bond 
length \cite{Yam90}. 
(b) Energy level scheme of Cu, \OP, and \OA. 
(c) Schematic picture of the screening effect: Once a charge is moved 
from one Cu-ion to another (at the energy cost $U$, initially), the apical 
oxygen orbitals experience an electric field (arrows) of excited charges 
and are polarized. An energy gain from the polarization process reduces 
a virtual charge excitation energy $U$. 
}
\label{fig1}
\end{figure}
%%%%%%%%%%%%%%%

Calculating the expectation value 
$\langle c_{-k\downarrow}c_{k\uparrow}\rangle$ in the model (1), 
we find the following gap equation:  
\begin{equation} 
1=g\sum_k\sum_{\pm} 
\frac{Z_{\pm}}{2E_{\pm}} \tanh \frac{E_{\pm}}{2T} \; |\gamma_k^d|^2. 
\label{gapeq}
\end{equation}
The quasiparticle energies are given by 
$E_{\pm}=\frac{1}{\sqrt{2}}
[\epsilon_{k}^2+\Delta_k^2+\epsilon_{\rm A}^2+2\nu_k^2\pm R^2]^{1/2}$, where 
$R^2=\{(\epsilon_{\rm A}^2-\epsilon_{k}^2-\Delta_k^2)^2+4\nu_k^2
[(\epsilon_{k}+\epsilon_{\rm A})^2+\Delta_k^2]\}^{1/2}$. 
Eq.~(\ref{gapeq}) is composed of two parts where 
$Z_\pm = (E_\pm^2-\epsilon_{\rm A}^2)/(E_\pm^2-E_\mp^2)$ represent 
the spectral weights of the pairing-active $x^2$-$y^2$ orbital on the two 
bands $E_{\pm}$. We note that $Z_{-}$ (which is the most relevant one) 
is reduced from its bare value ($=$1 at $\nu=0$) due to the orbital mixing. 
This has the effect of reducing effective value of $g$ in Eq.~(\ref{gapeq}).  

%%%%%%%%%%%%%%%%%%%%
\begin{figure}[t]
\includegraphics[height=10.5cm]{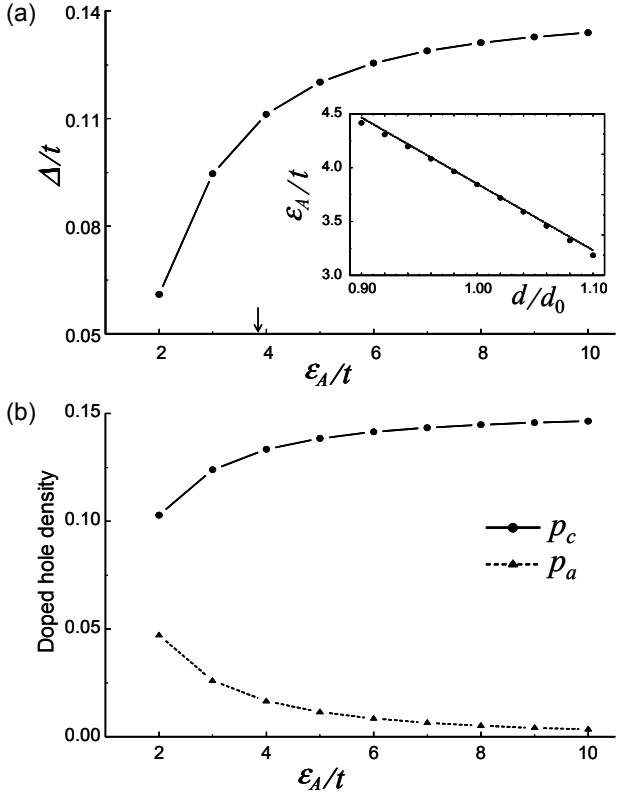}
\caption{
(a) The pairing gap $\Delta$ as a function of the energy-level 
separation $\epsilon_{\rm A}$ between the apical orbital $a$ and the 
$x^{\rm 2}$-$y^{\rm 2}$ band. Arrow indicates $\epsilon_{\rm A}$ at $d=d_0$.
Inset: $\epsilon_{\rm A}$ vs. Cu-\OA~distance $d$. 
The data are fitted by a linear relation 
$\epsilon_{\rm A}(d)/t =3.85[1-1.6 (d/d_0-1)]$. 
(b) Distribution of doped holes among the $x^2$-$y^2$ and the 
$a$ bands, denoted by $p_{\rm c}$ and $p_{\rm a}$, respectively. 
Total density of holes is fixed to $p_{\rm c}+p_{\rm a}$=0.15.
}
\label{fig2}
\end{figure}
%%%%%%%%%%%%%%%%%%

First, we regard $\epsilon_A$ as a free parameter and consider how its 
variation affects the gap. Solving Eq.~(\ref{gapeq}) at $g/4t=0.9$ and 
$T=0$, we obtained a sizable variation of the gap as a function of 
$\epsilon_A$ as shown in Fig.~2(a). This is due to the covalency effect 
that reduces the spectral weight of the $x^{\rm 2}$-$y^{\rm 2}$ states 
near the Fermi-level, by transferring it to the higher energy apical 
states. An amount of hole transferred into the apical level is rather 
small, of the order of several percent, see Fig.~2(b). However, it may 
be observed in the nuclear quadrupole resonance (NQR) which is sensitive 
to the hole concentration \cite{Zhe95}. 

Next, we consider how $\epsilon_{\rm A}$ is shifted by the structural
modulation. From the Madelung potential calculation, we obtained a linear 
relation 
$\epsilon_{\rm A}(d)/t =(\bar\epsilon_{\rm A}/t)[1-a\;(d/d_0-1)]$, 
with $\bar\epsilon_{\rm A}/t$=3.85 \cite{note2} and $a$=1.6. 
$d_0\simeq 2.4$ \AA~is an average Cu-\OA~distance \cite{Yam90}. 
The inset of Fig.~2(a) shows that the energy level separation 
$\epsilon_{\rm A}$ {\it increases} as the apical site \OA~comes 
closer to the Cu-ion. Consequently, the pairing gap is also increased 
[see Fig.~2(a)], since the strength of the hybridization is reduced. 
 
%%%%%%%%%%%%%%%%%%%%%%%%%
\begin{figure}[t]
\includegraphics[height=5.5cm]{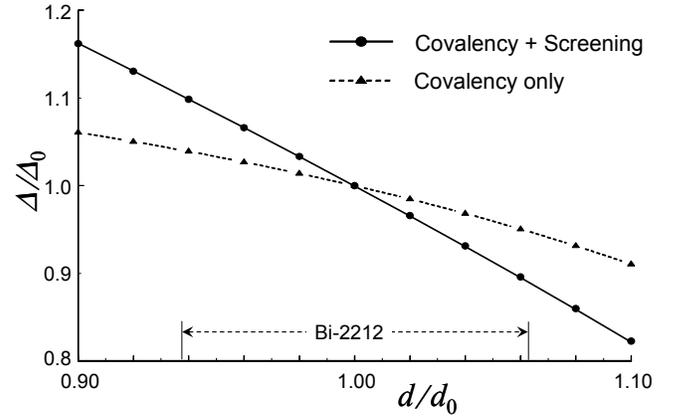}
\caption{
The gap $\Delta$ as a function of $d$, where 
$\Delta_0$$\equiv$$\Delta(d_0)$. The broken line shows the covalency 
effect only, i.e., the gap variation is solely due to modulation of 
$\epsilon_{\rm A}(d)$ at fixed $g/4t=0.9$. The solid line includes also 
the screening effect through the $d$-dependence of the coupling constant 
$g(d)$ (see text). In Bi-2212, Cu-\OA~distance varies within the range 
indicated. 
}
\label{fig3}
\end{figure}
%%%%%%%%%%%%%%%%%%%%%%%%%%%
Having obtained the relations $\Delta$ vs $\epsilon_{\rm A}$ and 
$\epsilon_{\rm A}$ vs $d$, we are now in position to show the 
gap $\Delta$ variations as a function of $d$ directly. 
The result is presented in Fig.~3 by the broken line, which shows 
that the $\pm$6\% change of $d$ leads to a sizable variation of $\Delta$. 

{\it Screening.}-- This effect is based on an observation \cite{Boe84,Bri95} 
that anion polarization renormalizes the energy of virtual charge 
excitations. This physics is relevant here since the \OA-contribution 
to the screening of in-plane interactions should strongly depend on $d$. 

Let us consider how $U$ and magnetic correlations within the CuO$_2$ planes 
are modified by apical \OA. In the $U$-excited intermediate state, both an 
unoccupied and doubly occupied Cu sites strongly polarize apical sites 
just above each Cu site, see Fig.~1(c). Apical oxygen obtains a dipole moment 
$p=\alpha F$ where $\alpha$ is the polarizability of O$^{-2}$ ion, and 
$F\simeq \rme/d^2$ is an electric field on oxygen induced by an extra 
charge (hole or electron) on Cu-site which is created by $U$-excitation. 
(We ignored dipole-field corrections to $F$ from further located ions 
for simplicity). Energy gain due to this polarization process in 
a virtual state reduces the $U$-excitation energy to 
$U_{\rm eff}(d)=U-E_{\rm pol}(d)$, where 
$E_{\rm pol}(d)=2(pF/2)=\alpha F^2\simeq \alpha(\rme/d^2)^2$  
is an interaction energy between the induced dipole moments $p$ on \OA~
and an excited hole (electron) at the unoccupied (doubly-occupied) Cu-sites. 
Since O$^{-2}$ ion has large polarizability, $\alpha\simeq $2 \AA$^3$ 
\cite{Sha93}, energy $E_{\rm pol}(d)$ is sizable: for 
the average Cu-\OA~distance $d_0\simeq 2.4$ \AA~we estimate 
$E_0\equiv E_{\rm pol}(d_0) \simeq$0.9 eV.  
It is important to realize that, due to the strong $d$-dependence 
of $E_{\rm pol}(d) \propto 1/d^4$, the effective repulsion $U_{\rm eff}$ 
becomes highly sensitive to the Cu-\OA~distance:  
$\delta U_{\rm eff}(d)=-\delta E_{\rm pol}(d)=4E_0~\delta d/d_0$.  
An immediate consequence of this observation is that the antiferromagnetic 
exchange interaction $J\simeq 4t^2/U_{\rm eff}$ between the Cu-spins 
obtains the same strong modulation as a function of $d$:   
$\delta J(d)/J_0 =-(4E_0/U_0)(d/d_0-1)$, where
$U_0\equiv U_{\rm eff}(d_0)$ and $J_0\equiv J(d_0)$. 
With the above estimate for $E_0$ and using a representative value 
$U_0\simeq$7 eV, we find that $\delta J(d)/J_0=-\beta (d/d_0-1)$ with 
$\beta \simeq 1/2$. Thus, in a modulated structure of Bi-2212, 
the $J$ value strongly increases as the Cu-\OA~distance $d$ decreases 
and vice versa, i.e., it shows the same anticorrelation effect with $d$ 
as the pairing gap $\Delta$ does in the experiment. 

Now, we assume that the superexchange driven magnetic correlations 
are essential for the pairing in cuprates as widely believed. 
Indeed, {\it e.g.}, in the $t-J$ model, $J$ plays a role of pairing potential 
same as $g$ in Eq.~(\ref{gapeq}). It is then natural to consider 
that the pairing potential $g$ in Eq.~(\ref{gapeq}) is modulated 
in the same functional form as $J$. We therefore implement a relation 
$g(d)=g[1-\beta(d/d_0-1)]$ with $\beta$=1/4 \cite{note3} and calculate 
the gap values from Eq.~(\ref{gapeq}) for different $d$. 
The obtained gap modulation is presented in Fig.~3 by the solid line. 
A combined action of the 
covalency and the screening effects can be summarized by a relation 
$\delta\Delta/\Delta_0 \approx -A\cdot\delta d/d_0$, with $A\simeq 1.6$. 

For a comparison of this result with experiment \cite{Sle08}, we notice that 
the measured gap $\Delta({\bf r})$ is in fact determined by a "coarse-grained" 
value $\tilde d({\bf r})\approx \langle d({\bf r}) \rangle_{\xi}$ 
of the actual Cu-\OA~distances. Such a coarse-graining of 
$\delta d \propto \cos(2\pi r_a/\lambda)$ \cite{Sle08} gives 
$\delta \tilde d({\bf r})\approx f\cdot \delta d({\bf r})$, i.e., the 
modulation "seen" by Cooper pairs is reduced by a factor $f$ that depends 
on the ratio of the coherence length $\xi$ and the supermodulation 
period $\lambda \approx$26 \AA~\cite{note4}. At $\xi\approx$20 \AA, we find 
$f\approx 0.43$. 
The above relation $\delta\Delta$ vs $\delta d$ reads then  
as $\delta\Delta/\Delta_0 \approx - \tilde A \cdot\delta d/d_0$, where 
$\tilde A \approx 0.7$. This gives $\approx$9\% total variation 
in $\Delta$ due to $\pm$6.25\% modulation of $d$, 
just as observed by Slezak {\it et al.}. 

In a broader context, it should be emphasized that while we are  
concerned here with the variations of apical oxygen position within 
{\it a given structure of a given material}, the effects 
discussed -- spectral weight transfer, and screening of effective $U$ -- 
are generic and relevant for the gap and $T_c$ variations among different 
cuprate families. However, many other things must be kept in mind 
when we compare different cuprates. In particular, apical oxygens 
may have a negative impact on $T_c$ by communicating a destructive 
effect of out-of-plane disorder to the CuO$_2$ planes \cite{Eis06,Liu06}. 
Anticorrelation between the energy gap and the distance to the apical 
oxygen observed by Slezak {\it et al.} in Bi-2212 implies that the effects 
we discussed here overcome the disorder related physics (which would result 
in a trend opposite to what observed). A key question is then how this 
competition is resolved in different cuprate families. To address this 
issue and better understand the $T_c$ trends in cuprates, our model has 
to be implemented by out-of-plane disorder effects.   

Finally, we argued that effective $U$ hence $J$ values are renormalized 
by \OA~and thus they become sensitive to the \OA-position. This 
inevitably turns the exchange interaction in Bi-2212 into inhomogeneous 
one in space. In other words, we expect that the strength of local spin 
correlations follow the lattice supermodulation. The resulting broad 
distribution of relaxation times could possibly be tested by the NMR/NQR 
experiments. In fact, the recent neutron scattering work has revealed 
an intrinsic broadening of the spin excitations in Bi-2212 \cite{Fau07}, 
an observation that seems natural in a light of our picture. 

To conclude, we discussed the physical origin of the relationship 
between the pairing energy gap and the atomic displacements in the 
supermodulated structure of Bi-2212. A covalent mixing of the $x^2$-$y^2$ 
orbital with apical $p$-level, and a screening of effective $U$ values 
via the polarization of apical oxygens are found to act cooperatively 
and modulate the pairing correlation as a function of the Cu-O$_{\rm A}$ 
distance. This leads to the spatial variations of the energy gap 
as observed in the experiment. 

We would like to thank J.C. Davis and B. Keimer for useful discussions. 
This work was supported by a Grand-in-Aid for Scientific Research on 
Priority Areas and the NAREGI Nanoscience Project from MEXT.  
M.M. was supported by a Grand-in-Aid for Young Scientists (B). 
G.Kh. thanks IFCAM at IMR, Tohoku University, and YKIS07, Yukawa 
International Program at YITP, for kind hospitality. The authors 
thank the Supercomputer Center, ISSP, University of Tokyo. 

%%%%%%%%%%%%%%%%%%%%%%%%%%%%%%


\begin{thebibliography}{99}
\bibitem{Oht91} 
Y. Ohta, T. Tohyama, and S. Maekawa,  
Phys. Rev. B {\bf 43}, 2968 (1991).

\bibitem{Mae04}
S. Maekawa {\em et al.}, 
{\em Physics of Transition Metal Oxides}, Springer Series in Solid State 
Sciences, vol. 144 (Springer-Verlag, Berlin, 2004). 

\bibitem{Mat90}
H. Matsukawa and H. Fukuyama,
J. Phys. Soc. Jpn. {\bf 59}, 1723 (1990).

\bibitem{Pav01}
E. Pavarini {\em et al.}, 
%I. Dasgupta, T. Saha-Dasgupta, O. Jepsen, and O.K. Andersen,   
Phys. Rev. Lett. {\bf 87}, 047003 (2001). 

\bibitem{Eis06} 
H. Eisaki {\em et al.}, 
%N. Kaneko, D.L. Feng, A. Damascelli, P.K. Mang, 
%K.M. Shen, Z.-X. Shen, and M. Greven, 
Phys. Rev. B {\bf 69}, 064512 (2004).

\bibitem{Liu06}
Q.Q. Liu {\em et al.}, 
%H. Yang, X.M. Qin, Y. Yu, L.X. Yang, F.Y. Li, R.C. Yu, C.Q. Jin, 
%and S. Uchida,
Phys. Rev. B {\bf 74}, 100506 (2006).

\bibitem{Mc05}
K. McElroy {\em et al.}, 
%Jinho Lee, J.A. Slezak, D.-H. Lee, H. Eisaki, S. Uchida, J.C. Davis,
Science {\bf 309}, 1048 (2005). 

\bibitem{Yam90} 
A. Yamamoto {\em et al.},
%M. Onoda, E. Takayama-Muromachi, F. Izumi, T. Ishigaki, and H. Asano,   
Phys. Rev. B {\bf 42}, 4228 (1990).

\bibitem{Sle07}  
J.A. Slezak, 
Ph.D. thesis, Cornell University, Ithaca, 2007. 

\bibitem{Sle08} 
J.A. Slezak {\em et al.}, 
%Jinho Lee, M. Wang, K. McElroy, K. Fujita, B.M. Andersen, 
%P.J. Hirschfeld, H. Eisaki, S. Uchida, and J.C. Davis, 
Proc. Natl. Acad. Sci. USA {\bf 105}, 3203 (2008). 

\bibitem{Nor08} 
M.R. Norman,  
Proc. Natl. Acad. Sci. USA {\bf 105}, 3173 (2008). 

\bibitem{And07} 
B.M. Andersen, P.J. Hirschfeld, and J.A. Slezak,  
Phys. Rev. B {\bf 76}, 020507(R) (2007).

\bibitem{Yan07} 
K.-Y. Yang, T.M. Rice, and F.-C. Zhang,  
Phys. Rev. B {\bf 76}, 100501(R) (2007).

\bibitem{lines} See the lines in Fig.~4 of Ref.~\onlinecite{Pav01}. 

\bibitem{phonon} Concerning phonons, experiment shows that $B_{\rm 1g}$ 
phonon modes (which may support $d$-wave pairing) are hardly 
affected by supermodulation \cite{Sle07}, and, moreover, their isotope 
shift has no effect on the gap variations. 

\bibitem{Boe84} 
D.K.G. de Boer, C. Haas, and G.A. Sawatzky,  
Phys. Rev. B {\bf 29}, 4401 (1984).

\bibitem{Bri95} 
J. van den Brink {\em et al.}, 
%M.B.J. Meinders, J. Lorenzana, R. Eder, and G.A. Sawatzky,   
Phys. Rev. Lett. {\bf 75}, 4658 (1995).

\bibitem{note1} 
The Madelung potentials are calculated in the ionic model, and are 
further divided by a dielectric constant $\epsilon_H=5$ [19] 
accounting for the screening of ionic potentials on a phenomenological level.

\bibitem{Hwa07}
J. Hwang, T. Timusk, and G.D. Gu, 
J. Phys.: Condens. Matter {\bf 19}, 125208 (2007). 

\bibitem{note2}
The result $\bar\epsilon_{\rm A}$=3.85$t$ is consistent with the band 
structure studies \cite{Nov93} which find $\sim 1.5-2.0$ eV 
separation between \OA~and \OP~derived bands. 

\bibitem{Nov93}
D.L. Novikov and A.J. Freeman, Physica C {\bf 216}, 273 (1993); 
D.J. Singh and W.E. Pickett, {\it ibid.} {\bf 235-240}, 2113 (1994).

\bibitem{Zhe95}
G.-q. Zheng, Y. Kitaoka, K. Ishida, and K. Asayama,
J. Phys. Soc. Jpn. {\bf 64}, 2524 (1995).

\bibitem{Sha93} 
R.D. Shannon, J. Appl. Phys. {\bf 73}, 348 (1993). 
%N.W. Grimes and R.W. Grimes, J. Phys.: Condens. Matter {\bf 10}, 3029 (1998).

\bibitem{note3} 
We have chosen a conservative value $\beta=0.25$ because, apart from 
the $U-$process, $J$ is contributed also by so-called charge-transfer 
term $J_{\Delta}=4t^2/(\Delta_{pd}+U_p/2)$ where the excited state 
contains two holes on $O_P$ with a correlation energy $U_p$. 
Considering the Madelung potentials and screening of the $pd$-transfer 
energy $\Delta_{pd}$ by apical oxygens in a similar way as for 
the $U-$process, we find that $J_{\Delta}$ is less modulated than $J_U$, 
i.e. $\beta_{\Delta}\simeq$0.1. Thus, an estimate $\beta\sim 0.3$ follows 
from $U\sim \Delta_{pd}+U_p/2$. 

\bibitem{note4} Specifically, $f=(2/x)J_1(x)$, where $J_1$ is Bessel function
and $x=\pi\xi/\lambda$. $f=1$ ($f=0$) if $\lambda\gg\xi$ ($\lambda\ll\xi$). 

\bibitem{Fau07}
B. Fauque {\em et al.}, 
%Y. Sidis, L. Capogna, A. Ivanov, K. Hradil, C. Ulrich, A.I. Rykov, 
%B. Keimer, and P. Bourges,
Phys. Rev. B {\bf 76}, 214512 (2007).

\end{thebibliography}
\end{document}